\documentclass [12pt]{article}
\makeatletter
\def\section{\@startsection {section}{1}{\z@}{-3.5ex plus -1ex minus
 -.2ex}{2.3ex plus .2ex}{\large\bf}}
\def\subsection{\@startsection{subsection}{2}{\z@}{-3.25ex plus -1ex
minus -.2ex}{1.5ex plus .2ex}{\normalsize\bf}}
\makeatother
\makeatletter

\@addtoreset{equation}{section}

\makeatother

\newcommand{\be}{\begin{equation}}
\newcommand{\ee}{\end{equation}}
\newcommand{\bea}{\begin{eqnarray}}
\newcommand{\eea}{\end{eqnarray}}
\def\one{{\rm 1\kern -.9mm l}}

\begin{document}
\begin{titlepage}
\rightline{NORDITA-2002-78 HE}
\vskip 1.8cm
\centerline{\Large \bf Non conformal gauge theories from D branes
\footnote{Work partially supported by the European
Commission RTN programme HPRN-CT-2000-00131.}}
\vskip 1.4cm \centerline{\bf P. Di Vecchia} \vskip .8cm
\centerline{\sl $^a$ NORDITA, Blegdamsvej 17, DK-2100 Copenhagen
\O, Denmark} \centerline{\tt divecchia@nbivms.nbi.dk}
\begin{abstract}
We use fractional and wrapped branes to describe perturbative and
nonperturbative properties of the gauge theories living on their
worldvolume.
\end{abstract}
\end{titlepage}
\renewcommand{\thefootnote}{\arabic{footnote}}
\setcounter{footnote}{0} \setcounter{page}{1}
\section{Introduction}

One of the most important ideas developed in recent years has been the
one that goes under the name of gauge-gravity
correspondence. According to it one can either use the low-energy
dynamics of  branes to study the properties of the gauge theory
living on them or, if one knows the properties of the gauge theory living
on  a brane, one can deduce its low-energy dynamics. This idea is also
at the basis of the Maldacena conjecture that, by using it, has established a
complete equivalence between a gauge theory (${\cal{N}}=4$ super
Yang-Mills ) and a superstring (supergravity) theory (type IIB string
theory compactified on $AdS_5 \times S^5$). In this paper we want to
use the gauge-gravity correspondence for studying the properties of
less supersymmetric and non-conformal gauge theories. We will not try to
establish an exact duality between these gauge theories and some
superstring theory as in the case of the Maldacena conjecture, but we will
use classical supergravity solutions corresponding to fractional and
wrapped branes having supersymmetric non-conformal gauge theories
living on them in order to study their perturbative and nonperturbative
properties. In particular we will use the expression of the gauge
coupling constant and of the $\theta$ angle in terms of the
supergravity fields in order to compute them when a consistent
classical solution is found.

For both wrapped D5 and fractional D3 branes of the orbifold $C^2 /
Z_2$ the gauge coupling constant is given by:
\begin{equation}
\frac{1}{g_{YM}^{2}} = \tau_5 \frac{( 2 \pi \alpha')^2}{2} \int
d^{2} \xi {e}^{- (\phi - \phi_0)} \sqrt{\det{( G_{AB} +
B_{AB})}}~~,~~
\tau_5 = \frac{1}{g_s \sqrt{\alpha'} ( 2 \pi \alpha')^5}
\label{gauge986}
\end{equation}
In the case of  wrapped branes that we will consider in this paper
we have to put $B=0$, while for fractional D3 branes, that for the
sake of simplicity we
take those of the $Z_2$ orbifold having only one vanishing two cycle, we get:
\begin{equation}
\frac{1}{g_{YM}^{3}} =  \frac{\tau_5 (2 \pi \alpha')^2}{2}
\int_{{\cal{C}}_2} {e}^{- \phi} B_2 = \frac{1}{4 \pi g_s (2 \pi
\sqrt{\alpha'})^2} \int_{{\cal{C}}_2} {e}^{- \phi} B_2
\label{coupli98}
\end{equation}
Finally the $\theta$ angle both in the case of fractional D3
branes and wrapped D5 branes is given by:
\begin{equation}
\theta_{YM} = \tau_5 (2 \pi \alpha')^2  (2 \pi)^2
\int_{{\cal{C}}_2} ( C_2 + C_0  B_2)
\label{theta56}
\end{equation}

The paper is organized as follows. In the next section we will
consider the case of fractional branes, while in section 3 we will use
wrapped branes for studying the properties of the gauge theory living
on them.

\section{Fractional branes}

In this section we will consider fractional D3 and D7 branes of the
orbifolds $C^2 /Z_2$ and $C^3 /(Z_2 \times Z_2)$ in order to study the
properties of respectively ${\cal{N}}=2$ and ${\cal{N}}=1$
supersymmetric gauge theories. The orbifold group acts on the
directions $x^4, \dots x^9 $ transverse to the worldvolume of the
D3 brane where the gauge theory lives. In particular in the case of
the first orbifold the nontrivial generator $h$ of $Z_2$ acts
as~\footnote{We denote $z_1 = x^4 + i x^5$, $ z_2 = x^6+i x^7$
and $z_3 = x^8 + i x^9$} $z_2 , z_3 \rightarrow - z_2 , - z_3$
while in the case of the second orbifold the three nontrivial
generators act as follows on the transverse coordinates:
\begin{eqnarray}
h \times 1  \Rightarrow & z_1 \rightarrow z_1~,~  z_2 \rightarrow -
z_2 , & z_3 \rightarrow - z_3  \nonumber \\
1 \times h   \Rightarrow & z_1 \rightarrow - z_1~,~  z_2 \rightarrow
z_2 , & z_3 \rightarrow - z_3  \label{orb5} \\
h \times h   \Rightarrow & z_1 \rightarrow - z_1 ~,~ z_2 \rightarrow
- z_2 , & z_3 \rightarrow  z_3  \nonumber
\end{eqnarray}
They are both non compact orbifolds with respectively one and three
fixed  points at the origin corresponding to the point $z_2 , z_3 =0$
and to the three points $z_1 , z_2 =0$, $z_1 , z_3 =0$ and $z_2 , z_3
=0$. Each fixed point corresponds to a vanishing $2$-cycle. Fractional
Dp branes are D(p+2) branes wrapped on the vanishing two-cycle and
therefore are, unlike bulk branes, stuck at the orbifold fixed point.
By considering $N$ fractional D3 and $M$ ($2M$) fractional D7 branes of the
two previous orbifolds we are able to study  ${\cal{N}}=2$
(${\cal{N}}=1$) super QCD with $M$ hypermultiplets.
In order to do that we need to determine the classical solution
corresponding to the previous brane configuration. For the case of the
orbifold $C^2 /Z_2$ the complete classical solution has been found in
Ref.~\cite{d3d7}~\footnote{See also
Refs.~\cite{d3,polch,grana2,marco} and Ref.~\cite{REV} for a review on
fractional branes.}.
In the following we
write it explicitly for  a system of $N$ D3 fractional
branes with worldvolume along the directions $x^0, x^1 , x^2 , $ and $
x^3$ and $M$ D7 fractional branes containing the D3 branes and having
the remaining four worldvolume directions along the orbifolded ones. The
metric, the $5$-form field strenght, the axion  and the dilaton are
given by~\footnote{We
denote with $\alpha$ and $\beta$ the four directions corresponding to
the worldvolume of the fractional D3 brane, with $\ell$ and $m$ those
along the four orbifolded directions $x^6 , x^7, x^8 $ and $x^9$ and
with $i$ and $j$ the directions $x^4$ and $x^5$ that are transverse to
both the D3 and the D7 branes.}:
\begin{eqnarray}
ds^2 &=& H^{-1/2}\, \eta_{\alpha\beta}\,d x^\alpha dx^\beta
+ H^{1/2} \,\left(\delta_{\ell m}\,dx^\ell dx^m + {\rm e}^{-\phi} \delta_{ij}
 dx^i dx^j\right)  ~~,
\label{met48} \\
{\widetilde{F}}_{(5)} &=&  d
\left(H^{-1} \, dx^0 \wedge \dots \wedge dx^3 \right)+ {}^* d
\left(H^{-1} \, dx^0 \wedge \dots \wedge dx^3 \right) ~~,
\label{f5ans}
\end{eqnarray}
\begin{equation}
\tau \equiv C_0 + i {\rm e}^{- \phi} =
{\rm i}\left(1 -\, \frac{Ng_s}{2\pi}\, \log
\frac{z}{\epsilon}\right)~~,~~ z \equiv x^4 + i x^5 = \rho {\rm e}^{i\theta}
\label{tausol}
\end{equation}
where the warp factor $H$ is a function of all coordinates that
are transverse to the D3 brane ($x^4,\ldots x^9$). The twisted fields
are instead given by  $B_2 = \omega_2 b$, $C_2 = \omega_2 c$
where $\omega_2$ is the volume form corresponding to the vanishing
$2$-cycle and
\begin{equation}
b {\rm e}^{-\phi} = \frac{(2 \pi \sqrt{\alpha'})^2}{2}
\left[ 1 + \frac{2N-M}{\pi} g_s \log \frac{\rho}{\epsilon} \right]~~,~~
c + C_0 b = - 2 \pi \alpha' \theta g_s (2N-M)
\label{bc63}
\end{equation}
It can be seen that the previous solution has  a naked singularity of
the repulson type at short distances. But, on the other hand, if we
probe it with a brane probe approaching the stack of branes
corresponding to the classical solution from
infinity, it can also be seen that  the tension of the probe vanishes
at a certain distance from the stack of branes that is larger than
that of the
naked singularity. The point where the probe brane becomes tensionless
is called in the literature enhan{\c{c}}on~\cite{enhancon} and at
this point the
classical solution cannot be used anymore to describe the stack of
fractional branes.

Inserting  in eq.s (\ref{coupli98}) and (\ref{theta56}) the classical
solution we get the gauge coupling constant and the $\theta$
angle~\cite{d3d7} :
\begin{equation}
\frac{1}{g_{YM}^{2}} = \frac{1}{8 \pi g_s} + \frac{2N-M}{8 \pi^2} \log
\frac{\rho}{\epsilon}~~,~~
\theta_{YM} =- \theta (2N-M)
\label{thetaym76}
\end{equation}
Actually in the case of an ${\cal{N}}=2$ supersymmetric theory one
gets in the gauge multiplet also a complex scalar field $\Psi$. This
means that, when we derive the Yang-Mills action from the Born-Infeld
action we also get a contribution from the kinetic terms of the brane
coordinates $x^4$ and $x^5$ that are transverse to the brane and
transverse to the orbifolded ones. This implies that the complex scalar
field of the gauge supermultiplet is related to the coordinate $z$ of
supergravity through the following gauge-gravity relation
$\Psi \sim \frac{z}{2 \pi \alpha'}$.
This is a relation between a quantity of the gauge theory living on
the fractional D3 branes and the coordinate $z$ of  supergravity. This
identification allows one to obtain the gauge theory anomalies from
the supergravity background. In fact, since we know how the anomalous
scale and $U(1)$ transformations act on $\Psi$, from
the previous gauge-gravity relation we can deduce how they act on $z$, namely
\begin{equation}
\Psi \rightarrow s {\rm e}^{2i \alpha} \Psi \Longleftrightarrow z
\rightarrow s  {\rm e}^{2 i \alpha} z   \Rightarrow \rho \rightarrow s
\rho~~,~~  \theta \rightarrow \theta + 2 \alpha
\label{gau78}
\end{equation}
Those transformations  do not leave unchanged the supergravity background in
eq.s (\ref{bc63}) and, as a consequence, they
generate the anomalies of the
gauge theory living on the fractional D3 branes. Acting with those
transformations on eq.s (\ref{thetaym76}) we get:
\begin{equation}
\frac{1}{g_{YM}^{2}} \rightarrow \frac{1}{g_{YM}^{2}} +\frac{2N-M}{8
\pi^2} \log s~~,~~ \theta_{YM} \rightarrow \theta_{YM} - 2 \alpha (2N-M)
\label{tra38}
\end{equation}
The first equation implies that the $\beta$-function of ${\cal{N}}=2$
super QCD with $M$ hypermultiplets is given by:
\begin{equation}
\beta (g_{YM} ) = - \frac{2N-M}{16 \pi^2} g_{YM}^{3}
\label{beta34}
\end{equation}
while the second one reproduces the chiral $U(1)$
anomaly~\cite{KOW,ANOMA}.
In particular, if we choose $\alpha = \frac{2 \pi}{2(2N-M)}$, then
$\theta_{YM}$ is shifted by a multiple of $2 \pi$. Since $\theta_{YM}$
is periodic of $2 \pi$, this means that the subgroup $Z_{2(2N-M)}$ is
not anomalous in perfect agreement with gauge theory results.

Using eq.s (\ref{thetaym76}) it is easy to compute
the combination:
\begin{equation}
\tau_{YM} \equiv \frac{\theta_{YM}}{2 \pi} + i \frac{4
\pi}{g_{YM}^{2}} = i \frac{2N-M}{2 \pi} \log \frac{z}{ \rho_{e}}~~,~~
\rho_e = \epsilon {\rm e}^{\pi/(2N-M)g_s}
\label{tauym67}
\end{equation}
where $\rho_{\epsilon}$ is called in the literature the enhan{\c{c}}on
radius and corresponds in the gauge theory to the dimensional scale
$\Lambda$ generated by dimensional transmutation.
Eq. (\ref{tauym67})  reproduces the perturbative moduli space of
${\cal{N}}=2$ super QCD, but not the instanton corrections. This
corresponds to the fact that the classical solution is reliable for
large distances in supergravity corresponding to short distances in the
gauge theory, while it cannot be used below the enhan{\c{c}}on radius
where nonperturbative physics is expected to show up. This means that
in order to study nonperturbative effects in the gauge theory we need
to find a classical solution free from enhan{\c{c}}ons and naked
singularities. This will be done in the next section. Before doing
that let us first extend the previous results to ${\cal{N}}=1$ super QCD
that can be obtained as a particular case  of the general one studied
in Ref.~\cite{NAPOLI}. In this case only the asymptotic behaviour for
large distances of the  classical solution has been explicitly obtained
and this is sufficient for computing the gauge coupling constant and
the $\theta$ angle of ${\cal{N}}=1$ super QCD. As explained in
Ref.~\cite{NAPOLI},
together with $N$ fractional D3 branes of the same type, one must also
consider two kinds of $M$ fractional D7 branes in order to
avoid gauge anomalies and one gets the following expressions for the
gauge coupling constant and the $\theta$ angle ( $z_i = \rho_i {\rm
e}^{i \theta_i}$)~\cite{ANOMA,FERRO,NAPOLI}:
\begin{equation}
\frac{1}{g_{YM}^{2}} = \frac{1}{16 \pi g_s} + \frac{1}{8 \pi^2} \left(
N \sum_{i=1}^{3} \log \frac{\rho_i}{\epsilon} - M \log
\frac{\rho_1}{\epsilon} \right)~~,~~\theta_{YM} =- N \sum_{i=1}^{3}
\theta_i + M \theta_1
\label{thetae4}
\end{equation}
As explained in Ref.~\cite{FERRO} the anomalous scale and $U(1)$
transformations act on $z_i$ as $z_i \rightarrow s {\rm
e}^{i2\alpha/3} z_i$. This implies that the gauge parameters are
transformed as follows:
\begin{equation}
\frac{1}{g_{YM}^{2}} \rightarrow \frac{1}{g_{YM}^{2}} + \frac{3N-M}{8
\pi^2} \log s~,~\theta_{YM} \rightarrow \theta_{YM} - 2 \alpha (N -
\frac{M}{3} )
\label{ano89}
\end{equation}
that reproduce the anomalies of ${\cal{N}}=1$ super QCD. The
differences between the anomalies in the ${\cal{N}}=2$ (eq.(\ref{tra38}))
and ${\cal{N}}=1$ (eq.(\ref{ano89})) super QCD can be easily understood
in terms of the different structure of the two orbifold considered. If
we consider the two gauge coupling constants there is a factor
$\frac{3}{2}$ between the contributions coming from the pure gauge
part, while the contribution of the matter is the same. The factor
$3$ is a consequence of the fact that the orbifold $C^3
/ (Z_2 \times Z_2)$ has three sectors, while the factor $\frac{1}{2}$
follows from an additional factor $\frac{1}{2}$ in the orbifold
projection for the orbifold $C^3 /(z_2 \times Z_2)$ with respect to
the orbifold
$C^2 /Z_2 $. This explains the factor $\frac{3}{2}$ in the gauge field
contribution to the $\beta$-function. The matter part is the same
because in the orbifold $C^2 /Z_2$ we have only one kind of fractional
branes, while in the other orbifold, in order to cancel the gauge
anomaly~\cite{NAPOLI}, we need two kinds of fractional branes. This
factor $2$ cancels  the factor $\frac{1}{2}$  coming from
the orbifold projection.
Similar considerations can also be used to relate the two chiral anomalies.

In conclusion, by using the fractional branes we have reproduced the
one-loop perturbative behaviour of both ${\cal{N}}=1$ and
${\cal{N}}=2$ super QCD, but, because of the enhan{\c{c}}on and naked
singularities we are not able to enter the nonperturbative region in
the gauge theory corresponding to short distances in supergravity. In
order to do this we must find a classical solution free of singularities.
That is why in the next section we turn to wrapped branes.

\section{Running coupling constant from wrapped branes}

In this section we turn to the case of wrapped branes and in
particular we will focus on a D5 brane wrapped on $S^2$ whose
corresponding solution, found in Ref.~\cite{CV} in four dimensions,
was riinterpreted as a ten dimensional one corresponding to a wrapped
D5 brane  and  used in
Ref.~\cite{MN} for describing ${\cal{N}}=1$ super Yang-Mills. A more
detailed and pedagogical derivation of the classical solution is
presented in Ref.~\cite{DLM} where  the classical solution was used
for determining the running coupling constant of ${\cal{N}}=1$ super
Yang-Mills as a function of the renormalization group scale $\mu$. In
particular, inserting the classical solution in  eq.(\ref{gauge986}),
one can determine how the gauge coupling constant depends on the
distance from the branes. One gets:
\begin{equation}
\frac{4 \pi^2}{ N g_{YM}^2} = F(\rho)
\label{gym87}
\end{equation}
But in order to determine the behaviour of the gauge coupling constant as
a function of the renormalization scale $\mu$ one must also give
a relation between $\rho$ and $\mu$. This was obtained in Ref.~\cite{DLM}
by connecting a certain function of $\rho$, called in Ref.~\cite{DLM}
$a(\rho )$, to the gaugino condensate following the suggestion
of Ref.~\cite{MILANO}. The result was:
\begin{equation}
a(\rho) = \frac{2 \rho}{\sinh 2 \rho}=
\frac{\Lambda^3}{\mu^3}~~.
\label{holo1}
\end{equation}
The running coupling constant is determined once we fix the
function $F( \rho )$ that depends on the two-cycle on which we wrap the
$5$ brane. On the other hand it is important to stress that the gauge coupling
constant depends on the renormalization scheme chosen and
therefore two different choices of the two-cycle  can be interpreted
to correspond to two different renormalization schemes.
In Ref.~\cite{DLM} the brane was wrapped on the $S^2$
spanned by the coordinates ${\tilde{\theta}}$ and ${\tilde{\varphi}}$
having chosen the other coordinates $\psi, \theta '$ and $\phi$ at
constant values~\footnote{We use the notation of Ref.~\cite{DLM}.}.
This choice gave the following result:
\begin{equation}
F( \rho ) = \frac{1}{4}
E\left(\sqrt{\frac{Y(\rho)-1}{Y(\rho)}}\right)~~,~~
Y(\rho) = 4\rho\,\coth 2\rho -1
\label{elli87}
\end{equation}
where $E$ is the elliptic integral and $F$ behaves as $\rho$ for large
values of $\rho$. In Ref.~\cite{DLM}, by considering only the leading
asymptotic behaviour of eq. (\ref{elli87}) and
by combining it with eq.(\ref{holo1}), it was derived that the
$\beta$-function of ${\cal{N}}=1$ super Yang-Mills was exactly the
NSVZ $\beta$-function~\cite{NSVZ} plus non perturbative corrections due to
fractional instantons~\footnote{An extension to the noncommutative
case was done in Ref.~\cite{MPT}.}. This result was questioned in
Ref.~\cite{OS}
where it was shown that, if one also includes the first non leading
logarithmic correction, one gets an extra contribution to the
$\beta$-function that modifies the one derived in Ref.~\cite{NSVZ}
already at two-loop level.  Then, in order to recover the correct
two-loop behaviour, it was suggested in Ref.~\cite{OS} to add in
eq.(\ref{holo1})  an extra  function $ f( g_{YM} )$ of the coupling
constant that can be fixed by requiring agreement with the correct
two-loop result. Of course it turns out that  $f ( g_{YM})$ must be
singular at $g_{YM} \sim 0$ as the transformation that is needed in
going from the holomorphic to the wilsonian
$\beta$-function~\cite{SV}. But, if we are prepared to recover the
correct two-loop behaviour by simply changing the renormalization
scheme, in order to
obtain the NSVZ $\beta$-function one could change immediately the scheme of
renormalization by trading the elliptic integral with just its asymptotic
behaviour: $ \frac{1}{4} E (\rho ) \rightarrow \rho$ as was
done in practice in Ref.~\cite{DLM}. This way of thinking eliminates a problem
that seems to appear if we perform a gauge transformation on the
non-abelian gauge field of gauged supergravity. In fact, if one
performs a gauge transformation in such a way that the gauge field
is vanishing in the  deep infrared ($\rho =0$), one gets a function
$F( \rho)$ that is different from the one in eq.(\ref{gym87}).
One gets~\cite{OLESEN}
\begin{equation}
F (\rho )= {\rm e}^{2h} + \frac{1}{4} (a-1)^2 = \rho \tanh \rho
\label{newf78}
\end{equation}
that, when put in eq.(\ref{gym87}), gives a Landau-pole singularity at $\mu
= \Lambda$ unlike the function in eq.(\ref{elli87}) that gave a smooth
behaviour at $\rho=0$. This is, however, not a problem if one also
interprets the gauge transformation in supergravity as a change of
renormalization scheme in the gauge theory.

A natural and elegant way to get directly the SNVZ $\beta$-function without
having to change the renormalization scheme as was implicitly done in
Ref.~\cite{DLM}, is presented in Ref.~\cite{BM} and is based on the
proposal of  choosing the  same cycle used in
Ref.~\cite{DLM} if one uses the solution after having performed the
previous discussed gauge transformation or equivalently use the
original solution and integrate on any of the two following cycles:
${\tilde{\theta}}= - \theta , {\tilde{\varphi}} =- \phi, \psi=0 $ or
${\tilde{\theta}}=  \theta , {\tilde{\varphi}} =- \phi, \psi=\pi $. In
both cases one gets precisely the expression in
eq.(\ref{newf78})~\cite{OLESEN,BM}.
This means that the definition of the two-cycle depends on which gauge
we use for the gauge field of gauged supergravity and if one takes
into account these changes one gets always the same result for the
gauge coupling constant of the gauge theory living on the wrapped $5$
brane.

In conclusion if one follows the proposal of Ref.~\cite{BM} the two
equations that determine the running gauge
coupling  constant of ${\cal{N}}=1$ super Yang-Mills as a function of
the renormalization scale $\mu$ are the following:
\begin{equation}
\frac{4 \pi^2}{ N g_{YM}^2} = \rho \tanh \rho~~~~;~~~~
\frac{2 \rho}{\sinh 2 \rho}=
\frac{\Lambda^3}{\mu^3}
\label{fine34}
\end{equation}
It is easy to check that they imply the NSVZ $\beta$-function plus
corrections due to fractional instantons. In fact from the previous
two equations after some simple calculation one gets:
\begin{equation}
\frac{\partial g_{YM}}{\partial \log \frac{\mu}{\Lambda}} \equiv
\beta (g_{YM} ) = - \frac{3 N g_{YM}^{3}}{16 \pi^2}
\frac{1+ \frac{2 \rho}{\sinh 2 \rho}}{\coth^2 \rho - \frac{N
g^{2}_{YM}}{8 \pi^2} - \frac{1}{2 \sinh^2 \rho}}
\label{beta23}
\end{equation}
This equation is exact and should be used together with the first
equation in (\ref{fine34}) in order to get the $\beta$-function as a
function of $g_{YM}$. It does not seem  possible, however, to trade $\rho$
with $g_{YM}$ in an analytic way. It can  be done in the
ultraviolet where,
from the first equation in (\ref{fine34}), it can be seen that
 $\rho$ can be approximated with
$\rho = \frac{4 \pi^2}{N g_{YM}^{2}} \coth \frac{4 \pi^2}{N
g^{2}_{YM}}$ obtaining the following $\beta$-function:
\begin{equation}
\beta (g_{YM} ) = - \frac{3 N g_{YM}^{3}}{16 \pi^2}
\frac{1+ \frac{4 \pi^2}{N g_{YM}^{2}} \sinh^{-2} \frac{4 \pi^2}{N g_{YM}^{2}}}{
\coth^2 \frac{4 \pi^2}{N g_{YM}^{2}} - \frac{N
g^{2}_{YM}}{8 \pi^2} - \frac{1}{2} \sinh^{-2} \frac{4 \pi^2}{N g_{YM}^{2}} }
\label{betafin}
\end{equation}
that is equal to the NSVZ $\beta$-function plus nonperturbative
corrections due to fractional instantons.

{\bf Acknowledgement} We would like to thank M. Bertolini, E. Imeroni,
P. Merlatti, R. Marotta, P. Olesen, F. Pezzella and
F. Sannino for  useful discussions and specially A. Lerda for many
exchanges of views on the subject of this talk.

\end{document}